\documentclass[]{spie}  
\usepackage[]{graphicx}
\usepackage[]{amsmath}
\usepackage[]{amssymb}

\title{Antenna-coupled TES Bolometer Arrays for BICEP2/Keck and SPIDER}

\author{A.Orlando\supit{a},  R.W. Aikin\supit{a}, M. Amiri\supit{b}, J.J. Bock\supit{ca}, J.A. Bonetti\supit{c}, J.A. Brevik\supit{a}, B. Burger\supit{b}, G. Chattopadthyay\supit{c}, P.K. Day\supit{c}, J.P. Filippini\supit{a}, S.R. Golwala\supit{a}, M. Halpern\supit{b}, M. Hasselfield\supit{b}, G. C. Hilton\supit{d}, K. D. Irwin\supit{d}, M. Kenyon\supit{c}, J.M. Kovac\supit{e}, C.L. Kuo\supit{fg}, A.E. Lange\supit{a}, H.G. LeDuc\supit{c}, N. Llombart\supit{c}, H.T. Nguyen\supit{c}, R.W. Ogburn\supit{afg}, C. D. Reintsema\supit{d}, M.C. Runyan\supit{a}, Z. Staniszewski\supit{c}, R. Sudiwala\supit{ah}, G. Teply\supit{a}, A. R. Trangsrud\supit{a}, A.D. Turner\supit{c} and P. Wilson\supit{c}
\skiplinehalf
\supit{a}Department of Physics, California Institute of Technology, 1200 E. California Blvd, Pasadena, CA 91125, USA; \\
\supit{b}Department of Physics and Astronomy, University of British Columbia, 6224 Agricultural Road, Vancouver, British Columbia, V6T 1Z1, Canada;\\
\supit{c}Jet Propulsion Laboratory, 4800 Oak Grove Dr, Pasadena, CA 91109, USA;\\
\supit{d}NIST Quantum Devices Group, 325 Broadway, Boulder, CO 80305, USA;\\
\supit{e}Harvard-Smithsonian Center for Astrophysics, 60 Garden Street, Cambridge, MA 02138, USA;\\
\supit{f}Department of Physics, Stanford University, 382 Via Pueblo Mall, Palo Alto, CA 94305, USA;\\
\supit{g}Kavli Institute for Particle Astrophysics and Cosmology, Sand Hill Road 2575, Menlo Park, CA 94025, USA;\\
\supit{h}School of Physics and Astronomy, Cardiff University, The Parade, Cardiff, CF24 3AA, UK.
}

\authorinfo{Corresponding author: A. Orlando : E-mail: orlando@caltech.edu}

\begin{document}
  \maketitle

\begin{abstract}
BICEP2/Keck and SPIDER are cosmic microwave background (CMB) polarimeters targeting the B-mode polarization induced by primordial gravitational waves from inflation. They will be using planar arrays of polarization sensitive antenna-coupled TES bolometers, operating at frequencies between $90~\mathrm{GHz}$  and $220~\mathrm{GHz}$. At $150~\mathrm{GHz}$  each array consists of 64 polarimeters and four of these arrays are assembled together to make a focal plane, for a total of 256 dual-polarization elements (512 TES sensors). The detector arrays are integrated with a time-domain SQUID multiplexer developed at NIST and read out using the multi-channel electronics (MCE) developed at the University of British Columbia. Following our progress in improving detector parameters uniformity across the arrays and fabrication yield, our main effort has focused on improving detector arrays optical and noise performances, in order to produce science grade focal planes achieving target sensitivities. We report on changes in detector design implemented to optimize such performances and following focal plane arrays characterization. BICEP2 has deployed a first $150~\mathrm{GHz}$ science grade focal plane to the South Pole in December 2009. \end{abstract}


\section{INTRODUCTION}
\label{sec:intro}  

One of the primary science goals of cosmic microwave background (CMB) cosmology in the next decade is the degree-scale B-mode polarization induced by a gravitational wave background. A detection would not only confirm inflation, but it would also distinguish between models and constrain the physical processes causing it\cite{Dodelson}. Searching for B-mode polarization presents several challenges:  a large number of sensitive detectors are required, as well as wide frequency coverage for astrophysical foregrounds monitoring and excellent control of polarization systematics. \cite{CMBtaskrep}  BICEP2/Keck and SPIDER are experiment designed to measure the polarization of the CMB. SPIDER\cite{Jeff} is a balloon-borne experiment targeting the CMB polarization at large angular scales. BICEP2\cite{Walt}and The Keck Array\cite{Chris} will be observing from the South Pole, aiming to detect the signature of inflation on degree angular scales ($ \ell \sim 100 $), taking advantage of the long integration time available from ground to go extremely deep on $  \sim 2 \% $ of the sky that has minimum astrophysical foregrounds. BICEP2/Keck and SPIDER focal planes employ planar arrays of dual-polarization antenna-coupled TES bolometers operating at frequencies between  $ \sim 90~\mathrm{GHz} $ and  $ 220~\mathrm{GHz} $ .
At the LTD13 conference we reported our early progress in characterizing engineering focal plane arrays\cite{Orlando2009}  and SQUID multiplexed readout for BICEP2 and SPIDER and our progress in microfabrication\cite{Bonetti2009}. Having achieved consistently high fabrication yield and reproducible and uniform device parameters, we have since then focused on characterizing and optimizing focal plane optical properties and sensitivity.
\newline \indent The paper is organized as follows. In Section \ref{sec:FPU} we provide a description of detector arrays and focal plane architecture. In Section \ref{sec:design} we describe changes to detector design implemented to optimize optical and noise performance. Measured optical properties and arrays performance are presented in Section \ref{sec:performance}. Future development plans are outlined in Section \ref{sec:future}.

   \begin{figure}[t]
   \begin{center}
   \begin{tabular}{c}
  \includegraphics[height=9.0cm]{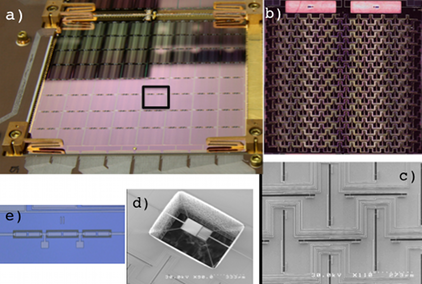}
   \end{tabular}
   \end{center}
   \caption[example]
   {\label{fig:polarimeter}
(a) Array of antenna-coupled polarimeters fabricated on a $4$-inch silicon wafer. Each small square (highlighted) is a complete polarimeter, for a total of $64$ at $150~\mathrm{GHz}$ ; (b) A polarimeter unit consists of a pair of co-locating orthogonal phased-array antennas, microstrip summing networks, microstrip filters and two TES bolometers, one for each linear polarization. The size of the polarimeter is $ \sim 7.5 ~\mathrm{mm} $ at $ 150~\mathrm{GHz} $.; (c) Details of the beam-forming antennas: orthogonal slot sub-antennas and summing network. The sub-antennas array format is  a $ 12 \times 12  $ cell for the polarimeter in (b); (d) SEM image of the thermally isolated silicon nitride island with the TES bolometer; (e) Microstrip bandpass filter.}
   \end{figure}

\section{Detector arrays and focal plane architecture} \label{sec:FPU}
At $150~\mathrm{GHz}$ each array consists of 64 polarimeters ($128$ TES detectors) fabricated on a 4-inch silicon wafer (see Fig.\ref{fig:polarimeter}a). A polarimeter unit (Fig.\ref{fig:polarimeter}b) consists of a pair of planar co-locating orthogonal phased-array antennas, microstrip summing networks, microstrip bandpass filters and two TES bolometers, one for each linear polarization\cite{Kuo2008} (called A and B). In this type of antenna-coupled device a planar array of slot antennas performs the function of beam collimation: the signals coming from the sub-antennas are coherently combined by a superconducting niobium microstrip summing network to form a beam. The beam width is approximately given by $ \lambda/d  $, where $ \lambda $ is the wavelength and $ d $ is the linear dimension of the antenna. Each polarimeter unit has two sets of orthogonal slots, readout by two independent microstrip networks, one for each polarization(see Fig\ref{fig:polarimeter}c). If the sub-antennas are arranged in a square grid pattern and fed uniformly the radiation pattern will exhibit minor sidelobes and four-fold symmetry. In refractor systems (such as BICEP2/Keck and SPIDER) the minor sidelobes are terminated at a cold Lyot stop. The array size is fixed by the wavelength and the desired $ \mathrm{FWHM}$:  for  $ \mathrm{FWHM}  \sim 14\,  ^{\circ} $ the sub-antennas array format is a $ 12 \times 12  $ cell and the size of the polarimeter is  $ \sim 7.5 ~\mathrm{mm} $ at $ 150~\mathrm{GHz} $.  Microstrip filters (Fig.\ref{fig:polarimeter}e) define both the upper and lower frequency cutoff of the science bands, with a chosen $  \sim  25\% $ fractional bandwidth, slightly smaller than the antennaÕs bandwidth.  After the bandpass filter the signal from the antenna is transmitted through the superconducting niobium microstrip and readout by a thermally isolated bolometer on a micromachined silicon nitride (SiN) island (see Fig.\ref{fig:polarimeter}d). The microstrip enters the thermally isolated island via a suspended SiN leg and terminates in a meandering resistive microstrip, where the electromagnetic energy is dissipated and detected by a TES bolometer, deposited on the same island. For more details on the SiN island design see Section \ref{sec:design}. Each TES bolometer consists of Ti ($ T_{c} \sim 520~\mathrm{mK} $) and Al ($ T_{c} \sim 1.34~\mathrm{K} $) connected in series. The Ti TES is used for science operations, the Al TES is used for optical characterization under laboratory loading conditions, where the Ti TES saturates.

   \begin{figure}[t]
   \begin{center}
   \begin{tabular}{c}
   \includegraphics[height=6.0cm]{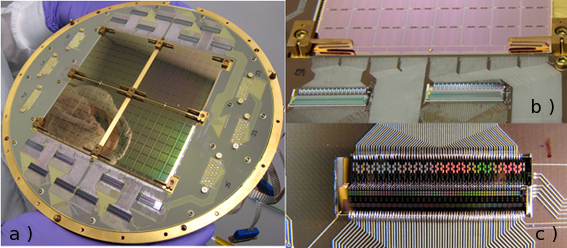}
      \end{tabular}
   \end{center}
   \caption[example]

   { \label{fig:FPU}
(a) Four arrays are assembled together to make a focal plane, for a total of $256$ polarimeters at $150~\mathrm{GHz}$. Each detector wafer, stacked with a  $ \lambda/4  $  quartz anti-reflection wafer (not visible), is  mounted on a gold plated OFHC copper plate using beryllium-copper spring clips fixed near the corners; (b) Close view of the spring clips holding a detector array. Each array is  connected to a printed circuit board via Al wire bonds. Visible are the aluminized traces on the printed circuit board, routing the signals from the detectors to $33$-element NIST SQUID MUX chips; (c) MUX/Nyquist chip, mounted on the printed circuit board using an intermediary ceramic carrier and wire bonded to the Al traces. 16 MUX/NYQ chips (visible in (a))  are used to readout all the detectors on the focal plane.}
   \end{figure}

\begin{figure}[t]
   \begin{center}
   \begin{tabular}{c}
   \includegraphics[height=5.5cm]{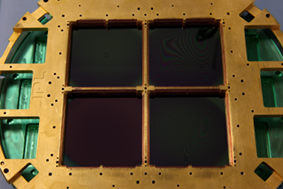}
   \includegraphics[height=5.5cm]{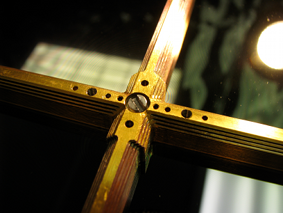}
   \end{tabular}
   \end{center}
   \caption[example]
   { \label{fig:corrugations}
\em Left \em: The radiation is coming through the quartz anti-reflection wafers, visible in the picture through the detector windows. \em Right \em:  Close view of the horizontal corrugations on the detector windows.}
   \end{figure}

Four arrays are assembled together to make a focal plane, for a total of 256 dual-polarization elements at $150~\mathrm{GHz}$ (see Fig.\ref{fig:FPU}a). The detector arrays are integrated with the SQUID time-domain multiplexer developed at NIST\cite{deKorte,Irwin2004}  using the following scheme : each detector wafer, stacked with a $ \lambda/4  $  quartz anti-reflection wafer, is mounted on a gold plated OFHC copper plate and connected to a printed circuit board, attached to the same stage, via Al wire bonds. Detector signals are routed via Al traces on the printed circuit board to 33-element NIST SQUID multiplexer (MUX) chips, attached to the same board (see Fig.\ref{fig:FPU}a/b). The first and second stage SQUIDs are on the same MUX chip, cooled at the same temperature as the detectors. We use 16 MUX chips to readout all the detectors on a focal plane. NIST ÒNyquistÓ inductor (NYQ) chips are used in conjunction with MUX chips to filter high frequency noise. The shunt resistors used to voltage bias\cite{Irwin95}  the TESs are fabricated on the NYQ chips at NIST. MUX/NYQ chips are mounted on the printed circuit board using an intermediary alumina carrier (see Fig.\ref{fig:FPU}c). Each detector wafer is held to the detector plate by beryllium-copper spring clips fixed near the corners (see Fig.\ref{fig:FPU}b). A niobium $ \lambda/4  $  backshort is mounted on top of the detector plate/printed circuit board assembly in Fig.\ref{fig:FPU}a, providing also shielding for MUX/NYQ chips. Radiation is coming from the opposite side, through the anti-reflection wafers from the detector arrays clear silicon side (see Figure \ref{fig:corrugations} (left)).

\noindent In order to achieve a better magnetic shielding, and not have to worry about signal induced by spinning the receiver in the earth's field, a different focal plane assembly has been designed for SPIDER \cite{marc}.

\noindent The $4$ detector arrays are readout on a $16$ columns by $32$ rows format using the multi-channel-electronics (MCE) \cite{elia2008} developed at the University of British Columbia (UBC). There are $32$ rows of detectors but $33$ rows of multiplexer channels due to the addition of a ``dark''  SQUID channel\cite{Irwin2004}, used to remove correlated low frequency noise in the amplifier chain. Columns are defined by the MUX/NYQ chips, while rows are defined by the $33$ first stage SQUIDs, each inductively coupled to a TES (with the exception of the ``dark'' SQUID).  We have one TES bias line for each NYQ chip, where the $32$ TESs of a given column are biased in series. Therefore we can bias all the detectors on the focal plane using $16$ bias lines, $4$ for each detector array.

Each array has $4$ ``dark '' detectors, identical to the polarization sensitive ones, but with the microstrip connection to the antenna broken. In addition to them, we mount one NTD thermistor on each detector wafer for accurate monitoring of thermal fluctuations. We actively servo the focal plane temperature and we can control the detector wafers temperature.
Particular effort has been made to avoid thermal gradients between the detector wafers and the copper plate. We use high conductivity z-cut crystal quartz for the anti-reflection wafers and we add gold wire bonds ($\sim 100$ per edge) between $3$ edges of each detector wafer and the gold plated copper plate to improve thermal conductivity. Original heat sink bond pads on the detector wafers have been replaced with a gold picture frame deposited directly on the silicon to increase the gold wires density.

As mentioned above, radiation is coming from the detector arrays clear silicon side through the anti-reflection wafers. Figure \ref{fig:corrugations} (left) shows the gold plated OFHC copper detector plate, with the $4$ detector windows and the $4$ anti-reflection wafers. Early engineering focal planes were suffering from poor optical performance near the edges of the detector windows,  showing low optical efficiencies and large A-B beam mismatch.  We simulated the effect of a metal edge on detectors A-B beams using a combination of CST Microwave Studio\footnote{http://www.cst.com}  and GRASP\footnote{http://www.ticra.com} simulation software.  The best of all the  geometries simulated was the case of horizontal corrugations on the detector windows  (Fig.\ref{fig:corrugations} (right)):  the beams are still displaced by the metal edge, but the effect is symmetric for the two polarizations, minimizing A-B beam mismatch. Simulations also showed that increasing the distance between antenna and detector plate edge also decreases the edge-pixel interaction. We designed a more compact detector array layout, reducing the spacing between pixel to pixel to increase the distance between pixel and detector window and minimize the interaction.
Both changes have been implemented on the BICEP2 science grade focal plane deployed to the South Pole, improving optical performance near the edges.

\section{Detector design optimization} \label{sec:design}

The main changes to the detector design consisted of: i) Redesigning the micromachined SiN island and suspended SiN legs and ii) Increasing the detectors heat capacity for BICEP2/Keck .
We also reduced the target thermal conductivity for BICEP2/Keck by a factor $ \sim 2 $ (to achieve $ G_{c} \sim 80~\mathrm{pW/K} $), taking advantage of the low and stable atmospheric loading at the South Pole.
   \begin{figure}[t]
   \begin{center}
   \begin{tabular}{c}
    \includegraphics[height=7cm]{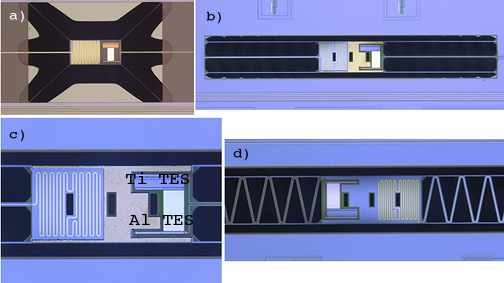}
   \end{tabular}
   \end{center}
   \caption[example]
   { \label{fig:island}
(a) Original silicon nitride island and legs design. The niobium ground plane running under the gold meandered resistor acts like a microstrip patch antenna, coupling radiation to the SiN island; (b) new island design (BICEP2): the distance between niobium ground plane and SiN island is only $25 \mu m$ and the niobium ground plane extends over the support legs; (c) detailed view of the new SiN island for a BICEP2 device: the meandered gold resistor is on the left, the TES bolometer on the right and a $2.5 \mu m$ thick layer of gold is deposited between them to increase the detector's heat capacity; $3$ holes had to be added to the island to avoid problems with the last fabrication process step (see text for details); (d) In order to achieve the lower $G$ required for SPIDER the SiN legs are similar, just longer, with a centered meander. No additional gold layer is deposited between the gold meandered resistor and the TES. } 
   \end{figure}

\subsection{Mitigating radiative coupling to the SiN island}
During early optical efficiency measurements on engineering grade focal plane the dark detectors on each detector array were showing a large response to changes in optical loading. At first we interpreted that as a thermal effect, due to an insufficient heat sinking of the wafers, caused by out-of-band radiation getting to the focal plane through our filter stack. After improving the heat sinking of the detector arrays (see Section \ref{sec:FPU} ) and using NTD thermistors and heaters to readout and control their temperature, we realized that the dark pixels response to optical radiation was not a thermal effect. Measurements performed using a chopped thermal source, in combination with low-pass edge filters\cite{Ade_LPE} with different frequency cutoffs and a thick-grille filter, showed that the dark pixels were detecting some amount of out of band radiation, with $\sim ~ 90\% $ of the ``blue leak'' at frequencies lower than $450~\mathrm{GHz}$ and about half of the ``blue leak'' at  $220 - 270 ~ \mathrm{GHz}$.  The only possible explanation was that radiation was coupling directly to the suspended SiN island. We simulated  the current SiN island design (shown in Fig.\ref{fig:island}a)) using CST Microwave Studio simulation software and it became evident that the niobium ground plane running underneath the gold meandered resistor on the SiN island was acting like a microstrip patch antenna, thus coupling radiation to the island. We simulated different possible solutions to this problem and the best was to bring the niobium ground plane as close as possible to the SiN island. The resulting design for BICEP2/Keck devices is shown in Fig.\ref{fig:island}b. All the SiN legs are now straight, the distance between SiN island and niobium ground plane has been reduced to $ 25~ \mu \mathrm{m} $ and the niobium ground plane  is deposited over the support SiN legs. The island size is  $310~\mu\mathrm{m} \times 150 ~\mu\mathrm{m}$ and it hasn't been modified. Great effort was required to solve fabrication problems with the new design. The final step in the fabrication process is  defining the thermally isolated SiN membranes. After patterning the SiN, bare silicon is exposed in areas except where island and legs will be. The exposed bare silicon is etched with a deep trench etcher, which cuts completely through the $500 ~\mu \mathrm{m}$  thick silicon. In the last step a $\mathrm{XeF_{2}}$ gas etch undercuts the silicon underneath island and legs.  With the new design we had to add $3$ holes to the island (see Fig.\ref{fig:island}c) to minimize the $\mathrm{XeF_{2}}$ etch time, allowing the Si underneath the island to etch away at the same time the legs clear, without overetching and undercutting the antenna.
In order to achieve the lower $G$ required for SPIDER the SiN legs are similar, just longer with a centered meander, shown in Fig.\ref{fig:island}d.
\newline
After changing the island design the residual coupling measured on dark pixels is  $\sim 1$-$2 ~\%$ of the signal on the `light' detectors at frequencies lower than $185~\mathrm{GHz}$.  We also measure a residual $\sim 0.3\%$ coupling on both dark and light pixels at frequencies higher than $185~\mathrm{GHz}$.

\subsection{Improving noise performance and stability}
Time-domain multiplexing is a powerful tool to readout a large number of detectors, but the limited sampling rate means that high frequency noise is aliased back into the signal band. The two main sources of high frequency noise are SQUID noise and detector noise. We find that aliasing increases the SQUID noise level by a factor  $ \sim  8$ - $10 $ ; however, this is a few times smaller than the detector noise on transition. High frequency intrinsic detector noise is filtered with a Nyquist inductor to reduce the amount of in-band aliasing. The inductance has to be large enough to avoid degradation due to aliasing, but still be small enough for detector bias stability\cite{Irwin2005}.  We have tested NYQ chips with inductances of   $ 0.5 ~ \mu H $ and  $ 1.35 ~ \mu H $ for both BICEP2 and SPIDER, measuring noise spectra at different bias points on the Ti transition. We found that an inductance of  $ 1.35 ~ \mu H $ is the best choice in order to not degrade sensitivity.  However, BICEP2 devices{'} time constants were fast enough that bias instabilities (electro-thermal oscillations) would start at bias points high ($> 0.7 ~ R_{n}$) on the superconducting transition, making almost impossible to find a stable operating point for all the detectors sharing the bias on each MUX column. SPIDER devices don't have the same problem : the lower thermal conductivity makes the thermal time constants slow enough that bias instabilities start only at bias points  $< 0.3 ~R_{n} $.

\noindent The easiest way to solve this problem and meet our sensitivity requirements for BICEP2/Keck was to design detectors with slower thermal time constants.  We tried increasing the detectors heat capacity by adding a thick layer of gold to the SiN island, between the TES and the gold meandered resistor. We fabricated test devices implementing few different SiN island geometries and gold layer thicknesses and we tested them with the SQUID multiplexer and $ 1.35 ~\mu H $ inductance NYQ chips. The best choice was to add a $ \sim  2.5 ~\mu m $ thick gold layer on the SiN island, covering all the surface available between the TES and  the gold meandered resistor, as shown in Fig.\ref{fig:island}c, thus increasing the detectors thermal time constant by a factor $ \sim 3 $. 
\newline More details on noise performance measured for BICEP2 devices after optimizing NYQ inductance and detectors time constant can be found in Brevik et al\cite{Justus} in these proceedings. 

\section{Arrays performance} \label{sec:performance}

The detector arrays are fabricated in the Micro Devices Laboratory at JPL.  Improvements in microfabrication process steps\cite{Bonetti2009} have made fabrication more and more reliable, reaching $ \sim  99 \% $ yield for the BICEP2 science focal plane. After fabrication the detector arrays are pre-screened at JPL performing extensive electrical checks, looking for antenna shorts to ground. Only arrays with $ >   96 \% $  yield are integrated with MUX/NYQ chips into science grade focal planes. Any further tests on SQUIDs and detectors are performed directly in BICEP2 and in the SPIDER test cryostat using the time-domain SQUID multiplexer and the MCE. MUX chips are usually pre-screened at NIST prior to assembly, performing SQUIDs critical current measurements at $4\mathrm{K}$. First stage SQUIDs on each row of the 16 MUX chips on the focal plane share the bias, so it is very important to select MUX chips with uniform critical currents per MUX row.
Using the MCE we can then easily characterize SQUIDs by measuring $V$-$\phi$ curves for each stage of the SQUID amplifier to find optimal operating points\cite{elia_squids}.


   \begin{figure}
   \begin{center}
   \begin{tabular}{c}
   \includegraphics[height=6.5cm]{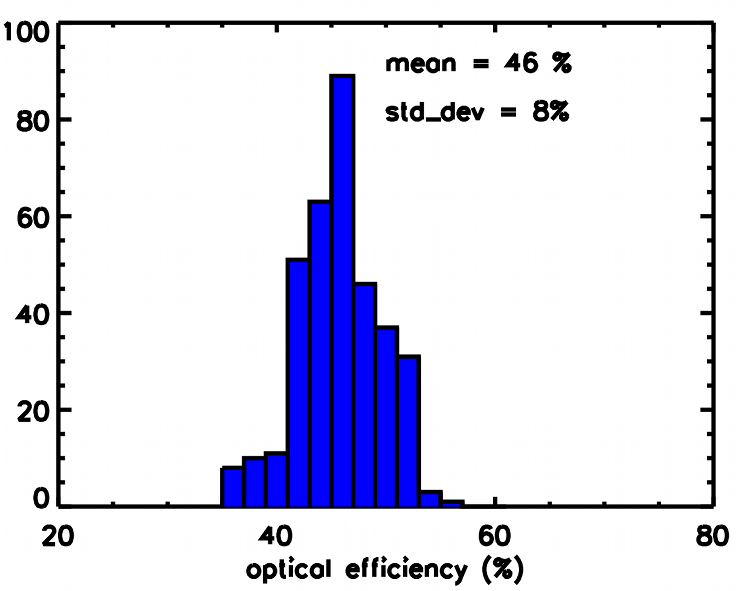}
   \end{tabular}
   \end{center}
   \caption[example]
   { \label{fig:opteff}
BICEP2 measured end-to-end optical efficiencies:  the average over the whole focal plane is $ \sim  46 \%$ ,  with a good uniformity.}
   \end{figure}
   
\subsection{Optical measurements}

Optical measurements have been routinely performed in BICEP2, under laboratory loading conditions, looking at thermal or coherent sources outside the dewar, and in the SPIDER test cryostat, under fliight-like loading conditions, using a helium cold load cryostat bolted to the top of the dewar, presenting a cold blackbody source to the instrument (the temperature can be elevated to $>  5K $ using a resistive heater)\cite{marc}.  BICEP2/Keck and SPIDER share the refractive optics design, but employ different filtering schemes, having different requirements in terms of total optical loading at the detectors. BICEP2 filter stack consists of two teflon blockers at $40\mathrm{K}$ and $100\mathrm{K}$ and a nylon blocker at $4\mathrm{K}$; a $8.3~\mathrm{cm}^{-1}$ low-pass edge metal mesh filter\cite{Ade_LPE} has been added to the nylon at $4\mathrm{K}$ before deployment. All these elements have been anti-reflection coated and optimized for transmission at $150~\mathrm{GHz}$ (for more details on the BICEP2 optics see Aikin et al in these proceedings\cite{Randol}). SPIDER{'}s filter stack consists of $2$ stacks of  $4$ metal-patterned mylar ``IR shaders''\cite{Carole} (at $20\mathrm{K}$ and $100\mathrm{K}$) to reduce the IR loading on the helium stage and anti-reflection coated low-pass edge metal mesh filters \cite{Ade_LPE} at $20\mathrm{K}$, $4\mathrm{K}$, as well as just above the focal plane at $1.6\mathrm{K}$. For more details on SPIDER{'}s instrument and optical performances see Runyan et al\cite{marc} in these proceedings.

Optical efficiency is measured by taking load curves for the Al TES at different optical loadings, corresponding (for BICEP2) to a beam-filling eccosorb cone at ambient ($\sim 300\mathrm{K}$) and liquid nitrogen ($\sim 77\mathrm{K}$) temperature. The difference in saturation power corresponds to the difference in absorbed optical power. The end-to-end optical efficiency is determined by comparing this change in saturation power to the expected optical power difference, given by the Rayleigh-Jeans equation for one polarization (for a single-moded detector) : $ \Delta P = k_{B} \Delta T  \nu  \Delta \nu$, where $k_{B}$ is the Boltzmann constant, $\nu$ is the band center frequency , $\Delta \nu$ is the spectral bandwidth and $\Delta T$ is the temperature difference between the two loads.  Figure \ref{fig:opteff} shows end-to-end optical efficiencies measured with the BICEP2 science grade focal plane deployed to the South Pole. The average over the $4$ detector arrays is $\sim  46 \% $, with a standard deviation of $ \sim  8 \%  $. The changes to detector arrays layout and detector plate described in Section \ref{sec:FPU}, as well as anti-reflection coating of all optical elements (lenses, filters and window), have greatly improved the end-to-end optical efficiencies measured in BICEP2 during early engineering test runs. \newline Typical end-to-end optical efficiencies measured for SPIDER at $150~\mathrm{GHz}$ are $ \sim 36 \% $.

   \begin{figure}
   \begin{center}
   \begin{tabular}{c}
  \includegraphics[height=5.0cm]{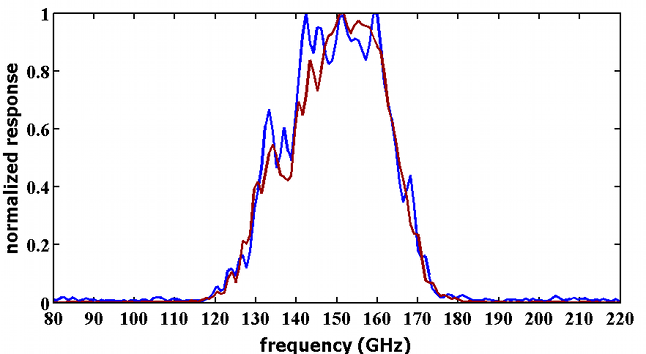}
   \end{tabular}
   \end{center}
   \caption[example]
   { \label{fig:spectra}
Typical spectra measured for a $ 150~\mathrm{GHz}$ pixel pair. The red and blue curves show well matched pass bands for the two orthogonally polarized devices on the same pixel, both independently normalized to unity.}
   \end{figure}

Filters spectra have been measured with a Martin-Puplett interferometer for $150~\mathrm{GHz}$ science grade arrays. The design center frequency is $148~\mathrm{GHz}$,  to better match the atmospheric transmission windows and avoid the $118~\mathrm{GHz}$  oxygen line and the $183 ~\mathrm{GHz}$ water line. Measured spectra (see Fig. 
 \ref{fig:spectra}) show that filter bands are very well matched between the two orthogonally polarized devices (A and B) in each pixel pair, with an average $\sim ~ 25\%$ bandwidth.

Polarization efficiencies have been measured for the BICEP2 focal plane at the South Pole, using a linearly polarized source in the far-field: the measured cross-polar leakage is  $  \leq  5 \times 10^{-3} $\cite{Randol}. 

\begin{table}[b]
\caption{Measured device parameters for the BICEP2 science focal plane: average $T_{c}$, $G_{c}$ (dark pixels only) and average saturation power ($P_{sat}$), measured under typical optical loading at the South Pole. Average measured parameters are listed for each detector array. In parenthesis next to the average $P_{sat}$ is the standard deviation.}
\label{tab:parameters}
\begin{center}
\begin{tabular}{|c|c|c|c|}
  \hline
 
  \rule[-1ex]{0pt}{3.5ex} Array & Gc (pW/K) & Tc (K) &$ P_{sat}$ @ Pole (pW) \\
  \hline
  \rule[-1ex]{0pt}{3.5ex} 1 & 132.0 & 0.525 & 14.8 ($12 \%$) \\
  \hline
  \rule[-1ex]{0pt}{3.5ex} 2 & 134.0 & 0.505 & 13.8 ($8 \%$) \\
  \hline
  \rule[-1ex]{0pt}{3.5ex} 3 & 79.0 & 0.517 & 7.2 ($12 \%$)\\
  \hline
  \rule[-1ex]{0pt}{3.5ex} 4 & 74.0 & 0.523 & 6.6 ($12 \%$) \\
  \hline
\end{tabular}
\end{center}
\end{table}

\subsection{Device parameters and arrays uniformity}

Having achieved reproducible and uniform device parameters\cite{Orlando2009}, we decided to optimize target thermal conductivities, aiming to  $G_{c}\sim 20~pW/K$ for SPIDER and  $G_{c}\sim 80~pW/K$ for BICEP2/Keck (at  $T_{c} \sim 510-520~mK$). Tests performed on few engineering detector arrays confirmed we can achieve the new targets  and easily optimize performances with optical loading.
For the BICEP2 science focal plane, having implemented all the design changes described in Section \ref{sec:design} shortly before deployment to the South Pole, we haven't characterized device parameters with a dedicated dark run. However, we have measured device parameters ($T_{c}$, $G_{c}$ and $\beta$) for the dark pixels on each detector array. The average values for each array are listed in Table \ref{tab:parameters}. We can see that $2$ arrays seem to have a thermal conductance at the transition temperature higher than target.

\begin{figure}
   \begin{center}
   \begin{tabular}{c}
   \includegraphics[height=6.5cm]{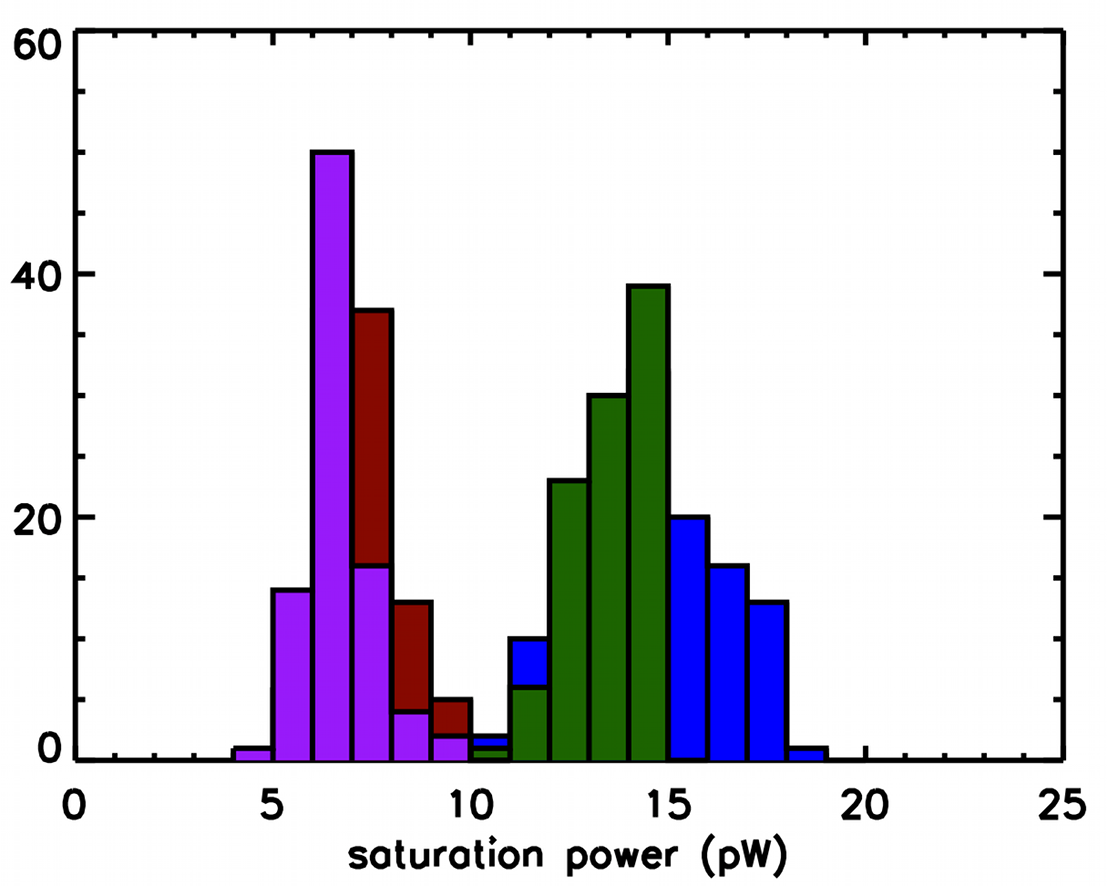}
    \includegraphics[height=6.5cm]{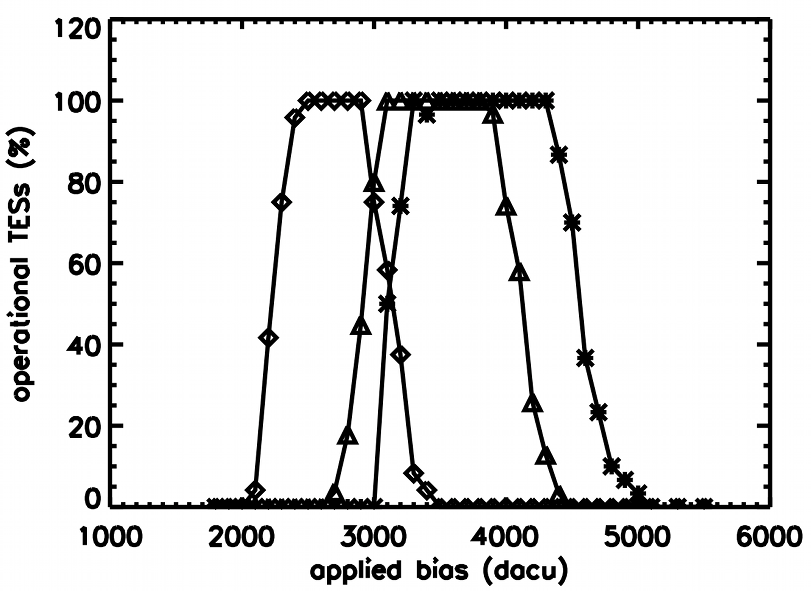}
   \end{tabular}
   \end{center}
   \caption[example]
   { \label{fig:Psat}
\em Left \em: Histograms of measured $P_{sat}$ (pW) for each array on the BICEP2 focal plane (at the South Pole). Measured values are consistent with measured $G_{c}$ listed in Table\ref{tab:parameters}. \em Right \em: fraction of TESs operational on transition at  at a given applied bias. We have one bias line for each MUX column, so we plot the fractions calculated for detectors on different MUX columns (for clarity only $3$ are shown). We can have $100\% $ detectors on each column operational at a wide range of applied bias and we can bias all the detectors on the focal plane by finding the optimum bias  on a column-by-column basis.}
   \end{figure}

Uniform device parameters and optical efficiencies are crucial to maximize the number of operational TESs under a given optical loading. We have measured BICEP2 detectors saturation power ($P_{sat}$) at the South Pole, under a typical optical loading (same as science observations) and at a focal plane temperature of $\sim ~ 0.280 ~K$. The measured values (calculated from detector load curves) are plotted in Fig. \ref{fig:Psat} (left) for each detector array. Average values and standard deviations are listed in Table \ref{tab:parameters}. We can see that uniformity over each array is really good ($\sim ~12\%$), but $P_{sat}$ has a bimodal distribution, in agreement with the device parameters measured for the dark pixels on the same arrays. 
\newline As described in Section \ref{sec:FPU}, on each focal plane we have $16$ TES bias lines, $4$ for each detector array, so only the $32$ detectors on each MUX column share the bias. Using detector parameters extracted from the load curves, we can calculate the number of operational TESs for each MUX column at a given applied TES bias (we consider a TES operational on transition if   $0.4~R_{n}~ < ~R_{tes}< 0.95~R_{n}$). The plot in Fig.\ref{fig:Psat} (right) shows the fraction of operational TESs as function of applied bias for $3$ MUX columns on $3$ different detector arrays (the other $12$ columns are not plotted for clarity, they show similar distributions). It is evident that we can have $100\%$ of the detectors on each column operational on transition for a wide range of applied bias; columns with different average $P_{sat}$ (shown in the plot) will just need a different applied bias to operate the TESs. Therefore during science observations at the South Pole we can operate all the working detectors on the BICEP2 focal plane by optimizing the applied bias on a column-by-column basis; thank to the good uniformity of device parameters and optical efficiencies we have a wide margin on the choice of applied bias. This is very important in order to optimize sensitivity, as the optimum bias selection is ultimately based on detectors noise properties and stability on transition, made more complicated by aliasing. More details on detectors noise and optimum bias points selection can be found in Brevik et al\cite{Justus} in these proceedings.

\section{Future work} \label{sec:future}
So far we have demonstrated good performances for science grade detector arrays at 150 GHz. BICEP2 has  deployed a first science focal plane to the South Pole in December 2009 and has been observing since February 2010. First results from BICEP2 observations of the main CMB field are reported in Ogburn et al\cite{Walt} in these proceedings.

Near field detector beam measurements (described in Aikin et al\cite{Randol} in these proceedings) show a ``beam-steering'' effect (beams are displaced from the center of the aperture) on edge pixels, one of the sources of pontentially large systematic contamination for CMB data. This effect could be caused by a spatial variation in the dielectric index of the planar phased antennas. We have started fabricating test arrays using different dielectrics for microstrip/antenna, in order to find the material with the smallest dielectric index variation and dielectric loss. Tests are currently underway and they will hopefully help improving optical performances of the detector arrays soon to be fabricated for the Keck Array and SPIDER.

Three $150~\mathrm{GHz}$ science focal planes for The Keck Array will be fabricated and tested in the next few months, in order to be deployed to the South Pole by December $2010$. Two $150~\mathrm{GHz}$ science focal planes for SPIDER will also be fabricated and tested, in preparation for an Antarctic flight in $2011$. Electrical tests of $90~\mathrm{GHz}$ engineering grade arrays for SPIDER have been successful, optical tests will start soon.

\acknowledgments    

We would like to thank the Gordon and Betty Moore Foundation, the National Aeronautics and Space Administration, the JPL Research and Technology Development Fund and the W.M. Keck Foundation. We also acknowledge NASA Postdoctoral Program support for Zak Staniszewski. 

\bibliography{report}   
\bibliographystyle{spiebib}   

\end{document}